# The effect of finite-temperature and anharmonic lattice dynamics on the thermal conductivity of $ZrS_2$ monolayer: self-consistent phonon calculations

Abhiyan Pandit[1,*] and Bothina Hamad[1,2,†]

[1]Physics Department, University of Arkansas, Fayetteville, AR 72701, USA

[2]Physics Department, The University of Jordan, Amman-11942, Jordan

**Abstract**

Two-dimensional (2D) $ZrS_2$ monolayer (ML) has emerged as a promising candidate for thermoelectric (TE) device applications due to its high TE figure of merit, which is mainly contributed by its inherently low lattice thermal conductivity. This work investigates the effect of the lattice anharmonicity driven by temperature-dependent phonon dispersions on thermal transport of $ZrS_2$ ML. The calculations are based on the self-consistent phonon (SCP) theory to calculate the thermodynamic parameters along with the lattice thermal conductivity. The higher-order (quartic) force constants were extracted by using an efficient compressive sensing lattice dynamics technique, which estimates the necessary data based on the emerging machine learning program as an alternative of computationally expensive density functional theory calculations. Resolve of the degeneracy and hardening of the vibrational frequencies of low-energy optical modes were predicted upon including the quartic anharmonicity. As compared to the conventional Boltzmann transport equation (BTE) approach, the lattice thermal conductivity of the optimized $ZrS_2$ ML unit cell within SCP + BTE approach is found to be significantly enhanced (e.g., by 21% at 300 K). This enhancement is due to the relatively lower value of phonon linewidth contributed by the anharmonic frequency renormalization included in the SCP theory. Mainly, the conventional BTE approach neglects the temperature dependence of the phonon frequencies due to the consideration of harmonic lattice dynamics and treats the normal process of three-phonon scattering incorrectly due to the use of quasi-particle lifetimes. These limitations are addressed in this work within the SCP + BTE approach, which signifies the validity and accuracy of this approach.

[*]*apandit@uark.edu,*

[†]*bothinah@uark.edu*

## 1. Introduction

It is well known that atoms in real crystals are not fixed at rigid lattice sites but vibrating from their equilibrium position. Lattice vibrations play a vital role in describing the thermal properties of crystalline solids by means of scattering in charge transports. The density functional theory (DFT) based on the standard harmonic approximation (HA) has been routine nowadays to study the phonon dispersion relations and several thermodynamical properties [1–4]. The HA is based on the second-order derivative of the Born-Oppenheimer (BO) energy surface around the ionic equilibrium, where the atomic displacements are considered sufficiently small compared to the interatomic distances. The HA is a useful technique to describe the elastic properties, phonon spectra and lattice vibrations of a crystal at 0 K. However, the lattice anharmonicity and temperature dependences of phonon, which play a key role to explain the lattice thermal conductivity (LTC) and thermal expansion in solids, cannot be accounted within the harmonic limit. In particular, the HA fails in the crystals that have significant anharmonic effects and for the materials that are dynamically unstable at 0 K.

The consistent explanation of equilibrium properties in crystals and the limited thermal conductivity of an insulating solid requires the presence of anharmonic terms [5], which include the cubic and higher-order terms in the energy expansion. Anharmonic effects are useful properties in the finite temperature phonon dispersion relation, phonon-phonon scattering rate, phonon lifetime, shift in phonon frequencies, and so on. These parameters play an important role in the phonon transport mechanism, which determines the LTC. The density functional perturbation theory (DFPT) [6–9] or the finite-displacement method [10] have been used to treat the anharmonicity. The finite-displacement approach uses the force-displacements data to extract the cubic, quartic and higher-order terms. However, this approach becomes computationally expensive as the range of the nearest neighboring atomic interaction and the order of anharmonicity increases. The DFPT, on the other hand, considers the anharmonic self-energies as a small perturbation of the harmonic terms, which is valid only when the anharmonic terms are sufficiently smaller than the harmonic term. The high temperature phases of the ferroelectric materials with the dynamical instability [11–15] and systems with light atoms (like hydrogen) [16–19] are some examples where the perturbation theory is not applicable. Therefore, a nonperturbative approach is needed to account the anharmonic effects for the emergent thermoelectric (TE) materials and hybrid perovskite solar cells [13,20–24].

Ab initio molecular dynamics (AIMD) simulations based approaches are generally used to account the anharmonic effects nonperturbatively [25–27]. However, the underlying problem here is in the Newtonian dynamics as its application is limited to the temperatures above Debye temperature. In addition, these AIMD based approaches are computationally expensive as well to obtain the converged phonon energies [28]. The temperature-dependent effective potential (TDEP) is one of the AIMD based approaches, which uses AIMD simulations and extracts the best possible harmonic or higher-order potential energy surfaces at finite temperatures [29]. But this approach, being an AIMD based, is limited when accounting for the zero-point vibration according to the aforementioned problem within the AIMD simulations. The self-consistent phonon (SCP) theory is another useful nonperturbative approach to include the lattice anharmonicity, which considers the quantum effect of phonons [30]. An efficient implementation of the SCP theory was developed recently [15,31], which inputs the higher-order anharmonic force constants obtained using the compressive sensing lattice dynamics (CSLD) technique [32].

Zirconium disulphide ($ZrS_2$) monolayer (ML) is a typical two-dimensional (2D) transition metal dichalcogenide that has been successfully synthesized [33–36]. Lowering the dimensionality and nano-structuring are reported to play major roles in suppressing the LTC that leads to an enhancement in the figure of merit ($ZT$) [37–41]. As compared to other similar 2D compounds, $ZrS_2$ ML is predicted to have a significantly low LTC upon using the conventional Boltzmann transport equation (BTE) within the relaxation time approximation (RTA), based on the harmonic phonon frequencies at 0 K [21]. In addition, the $ZrS_2$ ML has been predicted to have a higher $ZT$ value of 4.5 at 800 K, which is an important result for TE efficiency enhancement and device applications. However, the inadequate understanding of the lattice anharmonicity and its effect on the phonon dispersion and lattice thermal conductivity can be misleading in revealing the actual significance of $ZrS_2$ ML in thermoelectrics. One may wonder how does the lattice anharmonicity and finite temperatures affect the lattice dynamics and thermodynamic properties of the $ZrS_2$ ML? This important issue has not been examined to date, which is the stimulus of the present work. In this work we investigate the effect of finite-temperature anharmonic phonons on the thermal transport along with the LTC of the 2D $ZrS_2$ ML. We used an efficient first-principles method based on the SCP theory [15,31,42,43], which calculates the temperature-dependent phonon frequencies and LTC using the interatomic force constants (IFCs) within a supercell. The validity of this method was also well-demonstrated by carefully comparing the computed numerical results with experiments. We present brief details on the theoretical background of this work in section 2. The computational methods of the calculations are provided in section 3. The main results concerning the temperature dependence of the anharmonic phonons, lattice dynamics, thermodynamic parameters, and LTC are described in section 4. In section 5, the results are summarized by drawing final conclusions.

## 2.　　Theory

According to the BO approximation, the Hamiltonian ($H$) of an interacting nuclear system can be given by the sum of its kinetic energy operator ($T$) of ions and the potential energy ($U$) defined by the BO energy surface, $H = T + U$. The potential energy ($U$) of the system can be expressed as a Taylor expansion with respect to the atomic displacements as [15,44]:

$$U = U_0 + U_2 + U_3 + U_4 + \dots, \qquad (1)$$

where
$$U_n = \frac{1}{n!}\sum_{\{l,k,\mu\}} \Phi_{\mu_1,\dots,\mu_n}(l_1 k_1; \dots; l_n k_n) \times u_{\mu_1}(l_1 k_1), \dots, u_{\mu_n}(l_n k_n) \qquad (2)$$

is the $n$th-order ($n = 2, 3, \dots, \infty$) contribution to the potential energy, $u_\mu(lk)$ is the atomic displacement of the atom $k$ in the $l$th cell along the $\mu\,(= x, y, z)$ direction, and $\Phi_{\mu_1,\dots,\mu_n}(l_1 k_1; \dots; l_n k_n)$ represents the $n$th-order, or $n$-bodies, IFC, which is obtained by the $n$th-order derivative of the potential energy with respect to the atomic displacements. In Eq. (1) above, the linear term $U_1$ is neglected as the atomic forces are zero in equilibrium. In the HA, the expansion in Eq. (1) is truncated at the second order. The Hamiltonian here is transformed into Fourier space and then the harmonic phonon frequency $\omega$ is computed by constructing and diagonalizing the dynamical matrix.

The anharmonic contribution to the energy is generally expressed as a small perturbation $H'$ added to the harmonic Hamiltonian ($H_0 = T + U_0 + U_2$) as [15,31,45,46]:

$$H = H_0 + H' \approx H_0 + U_3 + U_4, \qquad (3)$$

where the higher order terms ($n > 4$) are omitted considering their contributions much smaller as compared to the third- and fourth-order terms. However, if the anharmonic terms are significant and comparable to the harmonic terms, they should be treated nonperturbatively. Using the SCP theory [15,31,47–49], which treats the anharmonic renormalization of the phonon frequencies nonperturbatively, the Hamiltonian in Eq. (3) can be re-written as:

$$H = \mathcal{H}_0 + (H_0 - \mathcal{H}_0 + U_3 + U_4) = \mathcal{H}_0 + \mathcal{H}', \quad (4)$$

where $\mathcal{H}_0$ ($= \frac{1}{2}\sum_q \hbar\Omega_q \Lambda_q \Lambda_q^\dagger$) is the effective harmonic Hamiltonian associated with the displacement operator $\Lambda_q$, $\Omega_q$ is the renormalized anharmonic phonon frequency, $\hbar$ is the reduced Planck's constant, and $q$ is the crystal momentum vector.

The free energy of the system can be obtained as the cumulant expansion of the term $\mathcal{H}'$ and the variational principle can be applied following the first-order SCP theory [31,43]. Then, the SCP equation can be derived as:

$$\Omega_q^2 = \omega_q^2 + 2\Omega_q I_q, \quad (5)$$

$$I_q = \sum_{q_1} \frac{\hbar \Phi(q;-q;q_1;-q_1)}{4\Omega_q \Omega_{q_1}} \frac{[2\tilde{n}_{q_1}+1]}{2}. \quad (6)$$

Where $\omega_q$ is the harmonic phonon frequency, $T$ is the absolute temperature, $\Phi(q;-q;q_1;-q_1)$ represent the fourth-order IFC, $\tilde{n}_{q_1} = \frac{1}{e^{\frac{\hbar\Omega_{q_1}}{k_B T}}-1}$ is the Bose-Einstein (BE) distribution function, $k_B$ is the Boltzmann constant, and $j$ is the phonon branch index for each $q$-point. Here, the anharmonic phonon frequencies ($\Omega_q$) are obtained by solving above Eqs. (5) and (6) self-consistently. During the iteration process of the SCP equation, the interpolation is performed using $q$ mesh within $q_1$ mesh for the inner loop.

## 3. Computational details

First-principles DFT calculations were performed by employing the VASP package [50,51], where the generalized gradient approximation (GGA) was used by applying the projector augmented wave method with the Perdew-Burke-Ernzerhof (PAW-PBE) functional [52]. The primitive unit cell of the 2D $ZrS_2$ ML was fully optimized within the self-consistent field loop convergence criteria of $10^{-6}$ eV. A cut-off energy of 500 eV and a Monkhorst-Pack $k$-mesh of 18 × 18 × 1 was used including the van der Waals interactions [53]. The atomic plane and its neighboring images were separated by a sufficiently large vacuum of 22 Å in the $z$-direction to avoid interactions between the mirror images during the DFT calculations. The electronic band structure calculations were also performed prior the lattice dynamics calculations.

A 5 × 5 × 1 supercell of ZrS$_2$ ML containing 75 atoms were used to extract the IFCs. To extract the harmonic IFCs, the finite-displacement approach was used. In this approach each atom was displaced from its equilibrium position by 0.01 Å including all possible nearest neighbor interactions. For the cubic IFCs, we considered up to the five-body nearest neighboring atomic interaction. The atomic forces on each of the displaced configurations were calculated, and then the cubic IFCs were extracted by using the ordinary least squares (OLS) fitting method with the harmonic IFCs as implemented in ALAMODE package [15,31,43,54]. A fitting error of 0.11 % was found here.

To estimate the IFC for the higher (quartic, or higher) order terms, the usual finite displacement method requires a large number (more than 5,000) of DFT calculations to get the atomic forces, which is computationally expensive. So, the quartic IFCs were extracted by using the CSLD technique [32], which is based on the machine learning programs as discussed and implemented in Refs. [15,31]. A 5 × 5 × 1 supercell of ZrS$_2$ ML was initially used to conduct AIMD simulations at 300 K for 14,000 MD steps with a time step of 1.5 fs. From the trajectory of AIMD simulations, 50 equally spaced atomic configurations were extracted. All the atoms in each of the atomic configurations were then displaced by 0.1 Å in random directions. The atomic forces were then calculated for these configurations using the DFT calculations. These calculated atomic forces were then used to estimate the quartic IFCs based on the least absolute shrinkage and selection operator (LASSO) technique [55], which solves the following equation:

$$\widetilde{\mathbf{\Phi}} = arg\ min_{\Phi}\ \|\mathbb{A}\mathbf{\Phi} - \mathcal{F}_{DFT}\|_2^2 + \alpha\ \|\mathbf{\Phi}\|_1. \qquad (7)$$

Here, $L_1$ regularization term is added to the least-squares method, $\mathbf{\Phi} = [\Phi_1, \Phi_2, \dots, \Phi_M]^T$ is the vector composed of $M$ linearly independent IFCs, $\mathcal{F}_{DFT}$ is the vector of atomic forces, and $\mathbb{A}$ is the matrix of atomic displacements. Moreover, $\alpha$ stands for the hyperparameter that controls the trade-off between the sparsity and accuracy of the model, whose optimal value was chosen from the cross-validation (CV) technique. The second order (harmonic) IFCs were fixed to the values obtained by OLS method and anharmonic terms were optimized during the LASSO regression step. The estimated quartic IFCs were used to calculate the anharmonic phonon frequency ($\Omega_q$) after solving Eqs. (5) and (6).

The thermodynamic parameters here can be evaluated as a function of the renormalized anharmonic frequency ($\Omega_q$). The specific heat capacity ($C_v$) can be calculated as:

$$C_V = \frac{k_B}{N_q} \sum_{q,j} \left(\frac{\hbar\Omega_{q_j}}{2k_BT}\right)^2 cosech^2\left(\frac{\hbar\Omega_{q_j}}{2k_BT}\right), \tag{8}$$

where $N_q$ is the number of $q$-points.

The LTC, which is an important parameter that contributes to the TE efficiency, can be computed generally by using the BTE within RTA as [31,56]:

$$\kappa_l^{BTE} = \frac{\hbar^2}{N_q V k_B T^2} \sum_q \omega_q^2 v_q \otimes v_q n_q (n_q + 1) \tau_q, \tag{9}$$

where $v_q = \frac{\partial \omega_q}{\partial q}$, $n_q$, and $\tau_q$ are the group velocity, BE distribution function and quasi-particle lifetime associated with the phonon frequency $\omega_q$, respectively; and $V$ is the volume of the unit cell. Here as the usual BTE approach treats the anharmonic effects perturbatively, where the temperature dependence of phonon frequencies and eigenvectors are neglected. Hence, the SCP theory is used to overcome this limitation, in which the LTC is modified as [15,31]:

$$\tilde{\kappa}_l^{SCP+BTE} = \frac{\hbar^2}{N_q V k_B T^2} \sum_q \Omega_q^2 \tilde{v}_q \otimes \tilde{v}_q \tilde{n}_q (\tilde{n}_q + 1) \tilde{\tau}_q, \tag{10}$$

where $\tilde{v}_q = \frac{\partial \Omega_q}{\partial q}$ and $\tilde{n}_q = n_q(\Omega_q)$ are usual terms associated with the renormalized phonon frequency $\Omega_q$. Here $\tilde{\tau}_q$ is the renormalized lifetime associated with the three-phonon scattering processes.

## 4. Results and discussions
### 4.1. Structural and electronic properties

The 2D ZrS$_2$ ML is found to crystallize in $1T$ structure (space group $p\bar{3}m1$) as shown in figure 1(a), where each Zr atom is octahedrally coordinated by six S atoms within the S-Zr-S sandwich layers. The unit cell of ZrS$_2$ ML is composed of three atoms (two Zr atoms and one S atom) with lattice parameters of a$_1$ = a$_2$ = 3.69 Å separated by angle of 120°. This result is in agreement with the previous experimental and theoretical results [21,33,34,57–59]. The electronic structure calculations predicted an indirect band gap where the valence band maximum (VBM) and conduction band minimum (CBM) are located at Γ and M points, respectively, as shown in figure 1(b). The band gap value is found to be 1.13 eV, without including the spin-orbit interaction (SOI), which remains almost the same (1.14 eV) upon including the SOI, see figure 1(b). These results are in agreement with previous studies [21,58,60,61]. The Brillouin zone with the selected high symmetry points for the electronic band structure calculation is shown in figure 1(c).

## 4.2. Anharmonic force constants from the compressive sensing technique

While the cubic terms are needed to perform the BTE calculations, the harmonic and quartic IFCs are necessary inputs to conduct the SCP calculations. The aforementioned LASSO technique was adopted to estimate the quartic IFCs, where the 50 displacement-force data sets obtained using AIMD are used. The predictive accuracy of LASSO regression was tested beforehand by using a four-fold cross-validation (CV) technique. The results of the CV are shown in figure 2, where figure 2(a) represents the relative error of the atomic forces, defined as a square root of $\| \mathbb{A}\widetilde{\Phi} - \mathcal{F}_{DFT} \|_2^2 / \| \mathcal{F}_{DFT} \|_2^2$, as a function of the hyperparameter $\alpha$. As the value of $\alpha$ decreases, the CV error decreases and reaches its minimum value at $\alpha = 4.96 \times 10^{-6}$ as indicated by the dotted vertical line in the figure. This value of $\alpha$ was chosen for the estimation of quartic IFCs because it is expected to give an accurate prediction for the data sets. With the chosen $\alpha$ value, we obtained a total of 9134 non-zero quartic IFCs as shown in figure 2(b), which represents about 55% of the total number of quartic IFCs with the rest of the physically irrelevant IFCs driven to be exactly zero.

## 4.3. Self-consistent phonons at finite temperature

Phonons are the vibrations resulting from the thermal energy of atoms or molecules within a system. The phonon dispersion curves and density of states (DOS) of $ZrS_2$ ML within the HA and SCP approach are shown in figure 3. Sufficiently converged anharmonic phonon frequencies ($\Omega_q$) were found with a $5 \times 5 \times 1$ $q$-grid points with respect to the intermediate $10 \times 10 \times 2$ $q_l$-grid points. The non-analytic correction was included by employing the Born effective charges and dielectric constants during SCP calculations. As the $ZrS_2$ unit cell has 3 atoms, there are a total of 9 phonon modes (3 acoustic and 6 optical), see figure 3(a). The lowest 3 phonon modes represent the acoustic modes, which are composed of in-plane longitudinal acoustic (LA), in-plane transverse acoustic (TA) modes, and out-of-plane flexural acoustic (ZA) mode. While LA and TA modes have a linear behavior near the $\Gamma$ point, the ZA mode exhibits a quadratic behavior, which is due to the 2D nature ML structure [37,62]. This behavior is consistent with the results obtained for other two-dimensional materials such as graphene[4,63–65], boron nitride [66,67], silicene [68], InX (Sn, Se, and Te)[69], and $MoSe_2$ [70]. The phonon frequencies of low-energy optical modes of $ZrS_2$ ML are found to increase within the SCP approach, which is due to the included quartic IFCs as shown in Eq. (5). The harmonic phonon frequency of the lower optical mode at Γ point is 145.36 cm$^{-1}$, which increases to 152.57 cm$^{-1}$ by including the quartic anharmonicity within the SCP approach at 0 K. The frequency further increases to 162.55 cm$^{-1}$ at 300 K temperature due to the temperature dependent term $I_q$ within the SCP calculations (see Eqs. (5) and (6)). The degeneracy of the optical modes 4 and 5 at the Γ point in the case of the harmonic phonon spectra is broken by the inclusion of the quartic anharmonicity within the SCP method, where the frequency of the optical mode 5 increases as seen in figure 3(a) with the corresponding phonon frequencies provided in table 1. This resolve of the degeneracy and hardening of the optical modes within SCP approach highlights the particular importance of the lattice anharmonicity for ZrS2, which is expected to affect the LTC later. The phonon DOS as a function of vibrational frequency are depicted in figure 3(b), where the higher peaks in the phonon DOS are found at the frequency region beyond 200 cm$^{-1}$. This can be attributed to the fact that the number of the lighter S atom, which contributes to the higher frequency, is double than that of the heavier Zr atom in the ZrS$_2$ ML unit cell.

## 4.4. Thermodynamic parameters

The specific heat capacity ($C_v$) is a crucial thermodynamic quantity, which directly contributes to the LTC ($\kappa_l$) as $\kappa_l \propto C_v$. The calculated $C_v$ as a function of temperature obtained within the SCP theory is shown in figure 4(a). The low values of $C_v$ at temperatures below 400 K indicate smaller contributions to LTC at low temperatures. The value of $C_v$ is found to increase as a function of the temperature until it converges to the classical limit of Dulong and Petit at higher temperatures. The mean-square displacement (MSD) is an important quantity that measures the deviation of atoms with respect to the equilibrium position in a system. The average MSD tensor of atom $k$ is computed as:

$$\langle u_\mu^2(k) \rangle = \frac{\hbar}{M_k N_q} \sum_{q,j} \frac{1}{\Omega_{q_j}} |e_\mu(k; q_j)|^2 (\tilde{n}_{q_j} + \frac{1}{2}), \tag{11}$$

where $e_\mu(k; q_j)$ is the corresponding atomic polarization.

There is an increase in the MSDs upon increasing the temperature, which is ascribed to the increased heating effect at higher temperatures, see figure 4(b). The increase in MSDs of thermal vibrations at higher temperature contributes to reduce the thermal transport. The calculated MSDs of S atoms are found to be higher than those of Zr atoms because of the higher atomic mass of Zr atoms and the inverse relation between atomic displacement and mass, see Eq. (11). This phenomena becomes more pronounced as the temperature increases. The total vibrational free energy is given by the sum of the free energy within the quasi-harmonic approximation (QHA) and the SCP correction term due to the anharmonicity within SCP method ($E_{Total} = E_{QHA} + E_{SCP}$). Although there is a negligible effect of SCP correction to the free energy within QHA at lower temperatures, the effect becomes more pronounced as the temperature increases as shown in figure 4(c). For instance, the SCP correction energy of -2.2 meV at 300 K decreases to -7.8 meV at 700 K. This decrease in total energy using the SCP correction indicates that the system is more stabilized by including the quartic anharmonicity. This fact also clearly highlights the importance of the anharmonic frequency renormalization on the thermal properties.

The phonon mode-dependent Grüneisen parameter ($\gamma_{qj}$) is another important dimensionless quantity that measures the anharmonic nature of the structure. This parameter is evaluated as the change of phonon frequency ($\omega_{qj}$ or $\Omega_q$) with respect to the change in volume (V): $\gamma_{qj} = -\frac{\partial(\log \omega_{qj})}{\partial(\log V)}$ by using the cubic IFCs. A positive value of $\gamma_{qj}$ indicates a decrease in the frequency of the phonon mode $q,j$ with the increase of volume. The calculated $\gamma_{qj}$ of ZrS$_2$ ML as a function of the phonon frequency is shown in figure 5(a), where the trend of change in $\gamma_{qj}$ values for both the SCP and harmonic lattice dynamics is found to be similar. The larger negative value of $\gamma_{qj}$ in the low frequency region (acoustic phonon modes) changes to small positive values at higher frequencies. This suggests larger phonon anharmonicity in the case of acoustic phonon modes. This nature closely affects the LTC via phonon life time ($\tau$) as $\tau_{qj}^{-1} \propto \gamma_{qj}^2$ according to the continuum theory [71,72].

The cumulative phonon group velocities ($v_g$) of ZrS$_2$ ML as a function of the phonon frequency is illustrated in figure 5(b). On average, group velocities of the acoustic phonon modes are found to be higher than those of the optical phonons. The average group velocities of the acoustic and optical phonons within the harmonic (SCP) lattice dynamics are 1.84 kms$^{-1}$ (1.83 kms$^{-1}$) and 1.1 kms$^{-1}$ (1.05 kms$^{-1}$), respectively. Hence, following the relation $\kappa_l \propto v_g$, the contribution of the acoustic phonon modes to the LTC should be larger than that of optical modes.

The phonon lifetime ($\tau$) is another significant quantity that is related to the LTC as $\kappa_l \propto \tau$ (see Eqs. (9) and (10)). It is calculated from the imaginary part of the anharmonic self-energy ($\Sigma_q(\omega_q)$) as $\tau_{q,anh}^{-1} = 2\Gamma_q^{anh} = 2Im\,\Sigma_q(\omega_q)$ [15,31,43], where $\Gamma_q^{anh}$ is the phonon linewidth.

Figure 5(c) shows the calculated $\tau$ of ZrS$_2$ ML at 300 K as a function of the phonon vibrational frequency by using the harmonic and SCP lattice dynamics, where the harmonic phonon frequency ($\omega_q$) is replaced by renormalized phonon frequency ($\Omega_q$) for the SCP lattice dynamics. The longer phonon lifetime is found in the low frequency range (acoustic modes) due to the low phonon-phonon scattering rate. This means the low frequency phonon modes have the major contribution to transport most of the heat in ZrS$_2$ ML, and consequently, a significant contribution to the LTC. As the obtained quasi-particle $\tau$ values by harmonic phonons are used in usual BTE approach, where the normal process of three-phonon scattering is incorrectly treated—this would result in $\kappa_l^{BTE} < \kappa_l^{actual}$ [31] as a consequence. The average value of $\tau$ using the SCP approach (2.24 ps) is higher than that of the harmonic phonons (1.59 ps) used in BTE approach, which is ascribed to the three-phonon scattering processes included within the SCP lattice dynamics.

### 4.5. Lattice thermal conductivity

The LTC ($\kappa_l$) spectrum as a function of the vibrational frequency and contributions of different phonon branches to $\kappa_l$ for the optimized unit cell of ZrS$_2$ ML are shown in figure 6. The low energy phonons below 160 cm$^{-1}$ are found to account for about 91% (94%) of the total $\kappa_l$ value within BTE (SCP + BTE) method at 300 K. The increase in cumulative $\kappa_l$ becomes negligible above the phonon frequency of 160 cm$^{-1}$ indicating the major contribution of the acoustic modes to $\kappa_l$. The significant decrease of $\kappa_l$ values above 160 cm$^{-1}$ is attributed to the increase in the available three-phonon scattering phase space (SPS), which describes the number of scattering channels available for a phonon. In ZrS$_2$ ML system, the three-phonon SPS at 300 K increases upon increasing the phonon frequency (above 160 cm$^{-1}$ as shown in figure S1, Supplemental Material), which corresponds to the subsequently lower $\kappa_l$ [73,74] thereby increasing the available scattering channels. The three phonon SPS values for absorption processes at 300 K reach to its minimum at higher frequencies (figure S1, Supplemental Material). In contrast, the SPS values for emission processes reach the peak at higher frequencies for both the harmonic and SCP methods. The contribution of different phonon branches to LTC is shown in figure 6(b). The low-energy acoustic phonon modes (branch 1 to 3) are found to have major contributions to the $\kappa_l$ values than the higher-energy optical modes (branch 4 to 9). In specific, the phonon branch 2 (branch 3) is found to have the highest contribution to $\kappa_l$ within BTE (SCP + BTE) approach.

Figure 7 shows the temperature dependence of LTC ($\kappa_l$) values of the DFT optimized unit cell of ZrS$_2$ ML calculated by using the conventional BTE and SCP + BTE approaches. The LTC calculations were performed using a sufficiently high 60 × 60 × 1 $q$-grid points with the neighboring images separated by a sufficiently large vacuum of 22 Å in the $z$-direction for both the BTE and SCP + BTE methods. The value of $\kappa_l$ is found to decrease with increasing temperature, in line with the standard relation $\kappa_l \propto T^{-1}$. This behavior is ascribed to the increase in the phonon linewidth ($\Gamma_q^{anh}$) (see figure S2, Supplemental Material) at higher temperatures, which is simply related to the increase in the BE distribution function with the temperature that leads to an increase in the scattering probability. The anisotropy in LTC values between the zigzag and armchair directions is appreciably enhanced within the SCP + BTE approach as illustrated in figure 7, which is attributed to the increase in the lattice anharmonicity upon the inclusion of the quartic IFCs. The calculated average $\kappa_l$ value of ZrS$_2$ ML at 300 K using the BTE method is 3.19 Wm$^{-1}$K$^{-1}$ (3.16 Wm$^{-1}$K$^{-1}$ along zigzag direction and 3.22 Wm$^{-1}$K$^{-1}$ along armchair direction), which is in agreement with the recently reported result of 3.29 Wm$^{-1}$K$^{-1}$ [21] obtained within the BTE method. The predicted $\kappa_l$ value of ZrS$_2$ ML in this work is found to be lower than that of bulk ZrS$_2$ along the in-plane direction, but higher as compared to that along the cross-plane direction [72], see table 2. It is remarkable that the calculated $\kappa_l$ values by SCP + BTE method are found to be relatively higher than those obtained by the conventional BTE method. At 300K, the average $\kappa_l$ value of ZrS$_2$ ML by using SCP + BTE approach is found to be 3.85 Wm$^{-1}$K$^{-1}$ (4.28 Wm$^{-1}$K$^{-1}$ along zigzag direction and 3.39 Wm$^{-1}$K$^{-1}$ along armchair direction), which is higher than the BTE value by 21%. A deeper insight into the origin of the difference in $\kappa_l$ can be obtained by comparing the calculated phonon linewidth ($\Gamma_q^{anh}$) within the BTE and SCP + BTE approaches (see figure S2, Supplemental Material), where $\kappa_l \propto (\Gamma_q^{anh})^{-1}$ as discussed above. The lower value of $\Gamma_q^{anh}$ within SCP + BTE approach (which uses SCP lattice dynamics) than that within the BTE approach (which uses harmonic lattice dynamics) leads to a higher LTC value for the SCP + BTE approach ($\kappa_l^{SCP+BTE} > \kappa_l^{BTE}$). These outcomes are in line with previous studies [15,31,42] that predicted the anharmonic lattice dynamics properties of cubic SrTiO$_3$ and SCF$_3$ using SCP + BTE approach, where the calculated results using SCP + BTE approach were found to agree well with the experimental results. Most crucially, the usual BTE approach neglects the temperature dependence of the phonon frequencies as it takes only the harmonic phonon frequencies into account. With these all facts, the $\kappa_l$ value calculated using the SCP + BTE approach is expected to be more accurate, and helps to predict reliable TE figure of merit value and energy conversion efficiency. The prediction should be validated by a future experimental study.

## 5. Conclusion

The lattice dynamical properties of 2D ZrS$_2$ ML are investigated using the SCP theory. The CSLD technique has proven to be a useful approach, as an alternative of the computationally expensive DFT calculations, to estimate the higher order anharmonic IFCs. The temperature-dependent phonon frequencies renormalized with the quartic anharmonicity are calculated nonperturbatively based on the SCP approach. The frequency renormalization phenomenon is found to be more pronounced at the low-energy optical modes. The LTC value obtained using the SCP + BTE approach is enhanced as compared to that predicted using the conventional BTE method. This is attributed to the relatively lower phonon linewidth due to the anharmonic phonon frequency renormalization phenomena included within the SCP theory. The conventional BTE approach i) neglects the temperature dependence of the phonon frequencies due to the consideration of harmonic lattice dynamics and ii) the normal process of three-phonon scattering is incorrectly treated due to the use of quasi-particle lifetimes. These limitations are addressed in this work within the SCP + BTE approach, which indicate the validity and accuracy of the approach. We expect that the present work not only correct the $\kappa_l$ value of ZrS$_2$ ML reported previously but also provides an insight to explain the effect of phonon anharmonicity on the lattice dynamics and thermodynamic properties. The effect of anharmonic frequency renormalization on the lattice dynamics of similar ultralow-LTC 2D materials followed by the experimental findings could be interesting future works regarding the TE efficiency measurement and device applications.

**Data availability statement**
The data that support the findings of this study are available upon reasonable request.

**Acknowledgements**
A. Pandit thanks T. Tadano for the useful discussions. All calculations were performed through Arkansas High Performance Computing Center at the University of Arkansas.

**Supplementary material**

Supplementary material for this article is available online at [**URL LINK**].

**Author contributions**

A. Pandit performed all the calculations and wrote the manuscript. B. Hamad supervised the work and edited the manuscript.

**Conflicts of interest**

The authors declare no competing conflict of interest.

# FIGURES

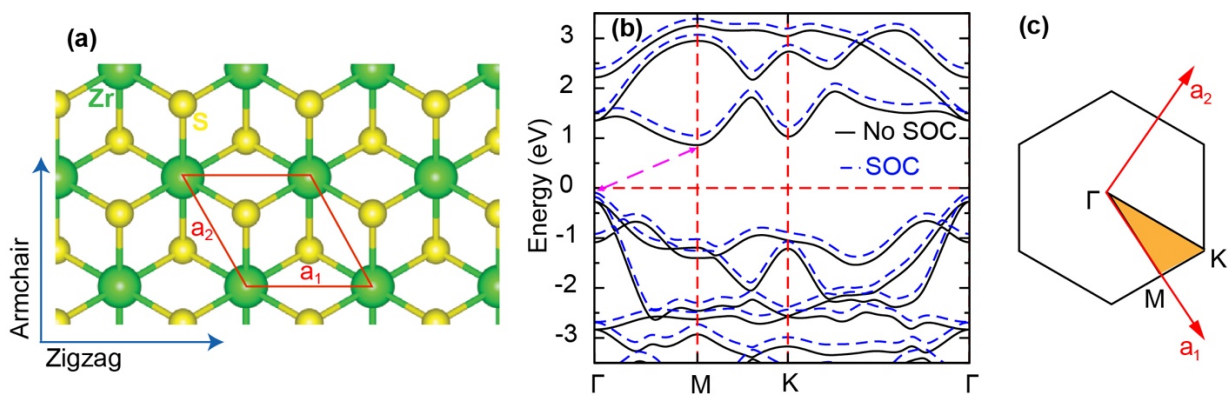

**Figure 1.** (Color online) (a) Crystal structure of $ZrS_2$ ML, the red rhombic box denotes the primitive unit cell. (b) The electronic band structure of $ZrS_2$ ML along the high-symmetry path $\Gamma$ – M – K – $\Gamma$ in the first Brillouin zone, the solid and dotted lines indicate the bands without and with spin-orbit interaction, respectively. (c) The Brillouin zone with labelled high-symmetry points.

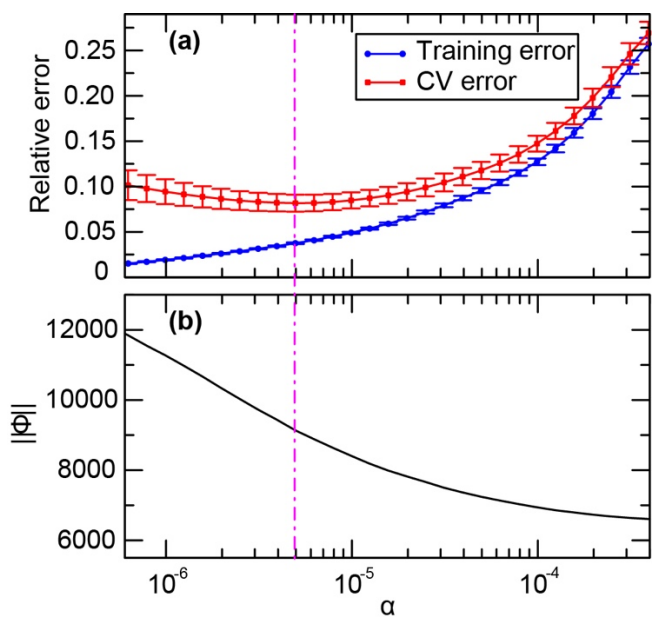

**Figure 2.** (Color online) (a) Relative errors in the atomic forces and (b) the number of non-zero quartic IFCs as a function of the hyperparameter ($\alpha$). The dotted vertical line represents the value of $\alpha$ at which the cross-validation error is minimum.

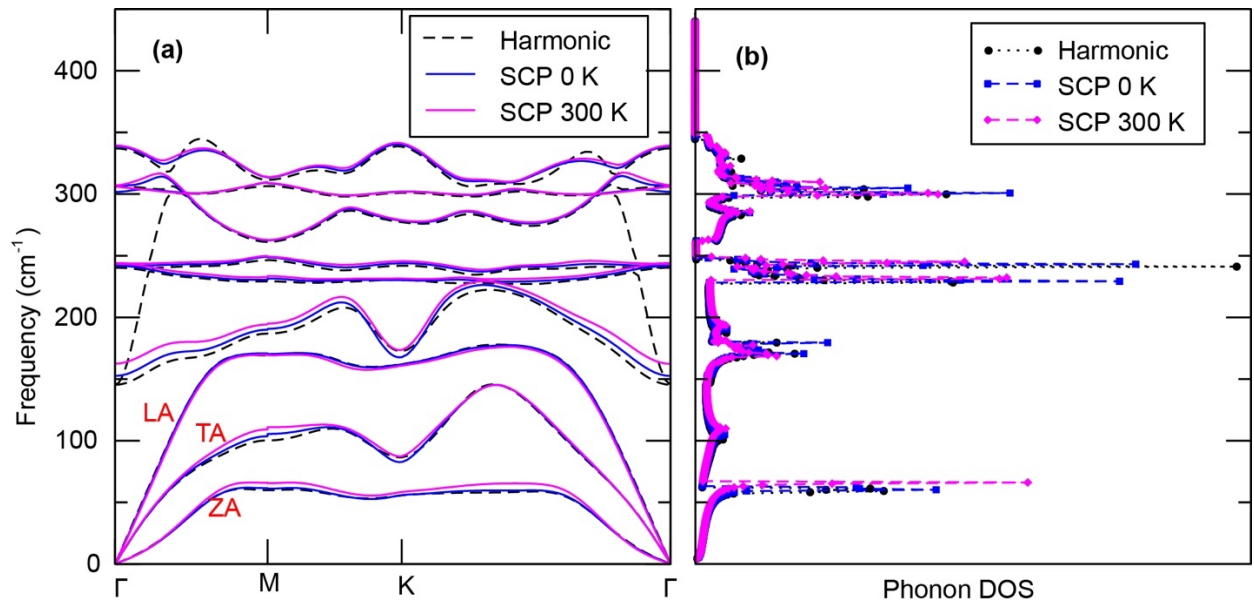

**Figure 3.** (Color online) (a) Temperature-dependent anharmonic phonon dispersion and (b) the phonon density of states (DOS) of the $ZrS_2$ ML. The black dotted lines indicate the harmonic lattice dynamics.

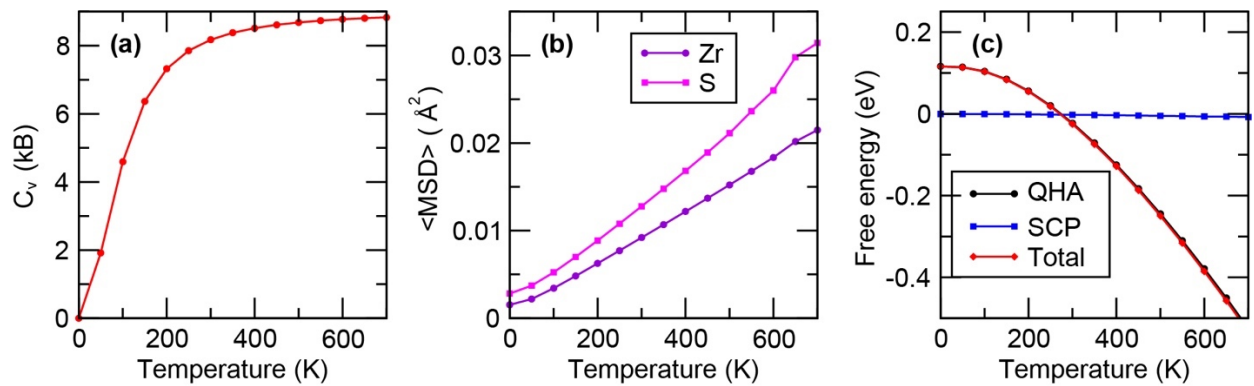

**Figure 4.** (Color online) Temperature dependence of (a) the specific heat capacity and (b) the root mean-square displacement of the Zr and S atoms, and (c) free energies within the QHA and SCP correction for the $ZrS_2$ ML.

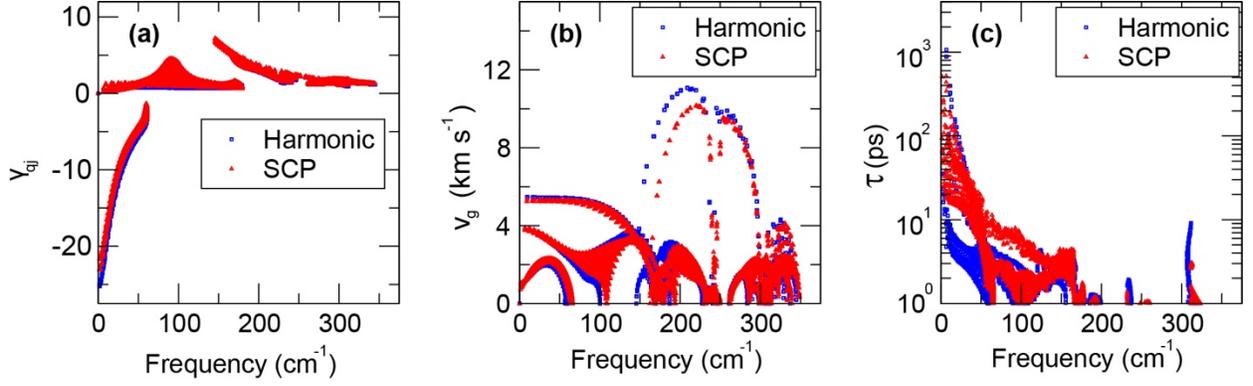

**Figure 5.** (Color online) (a) Grüneisen parameter, (b) the cumulative phonon group velocity, and (c) the phonon lifetime of ZrS$_2$ ML as a function of the phonon vibrational frequency obtained with the harmonic and SCP lattice dynamics.

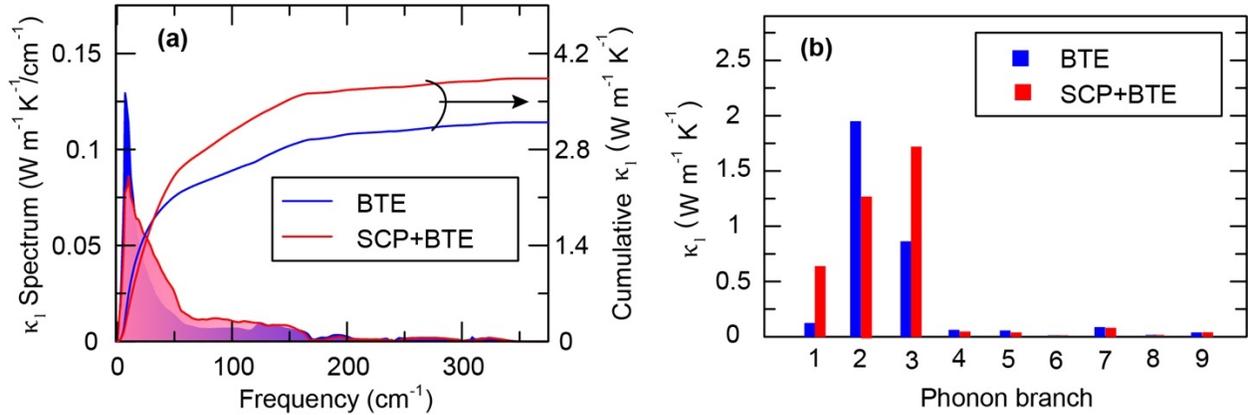

**Figure 6.** (Color online) (a) LTC ($\kappa_l$) spectrum and cumulative $\kappa_l$ as a function of the phonon frequency and (b) contributions of different phonon branches to the LTC of 2D ZrS$_2$ ML at 300 K obtained with the BTE and SCP + BTE methods. In (a), the curves with shaded region indicate the $\kappa_l$ spectrum and the curves without shade indicate cumulative $\kappa_l$.

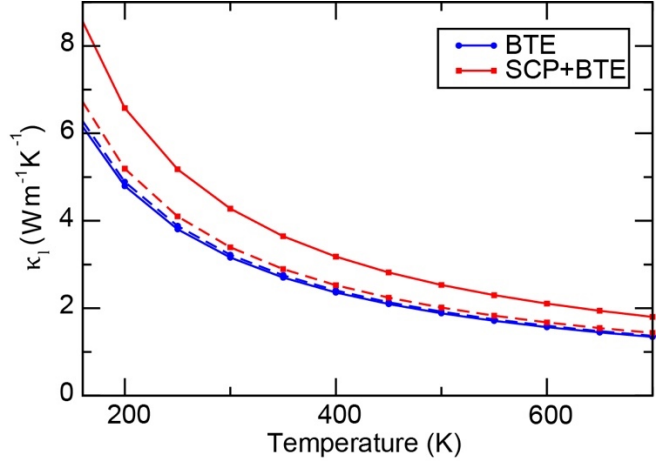

**Figure 7.** (Color online) LTC ($\kappa_l$) of the ZrS$_2$ ML as a function of temperature obtained with the BTE and SCP + BTE approach. The solid and dotted lines indicate the values along zigzag and armchair directions, respectively.

**TABLES**

**Table 1.** Phonon frequencies (cm$^{-1}$) of the lower optical modes of ZrS2 ML calculated using different approaches.

| Phonon modes | Harmonic | SCP at 0 K | SCP at 300 K |
|---|---|---|---|
| 4 | 145.36 | 152.57 | 162.55 |
| 5 | 145.38 | 230.55 | 243.45 |

**Table 2.** Calculated LTC of 2D ZrS$_2$ ML and its bulk system at 300 K.

| System | $\kappa_l$ (Wm$^{-1}$K$^{-1}$) at 300 K | References |
|---|---|---|
| ZrS$_2$ ML | 3.19 | This work (BTE) |
| | 3.85 | This work (SCP + BTE) |
| | 3.29 | Previous work (BTE) [21] |
| ZrS$_2$ Bulk | 8.5 (In plane) | Previous work (BTE) [72] |
| | 1.4 (Cross plane) | |